\documentclass[prd,nofootinbib,preprint]{revtex4}
\usepackage{graphicx}
\usepackage{epsfig}
\usepackage{makecell}
\usepackage{amsmath}
\usepackage{amsfonts}
\usepackage{amssymb}
\usepackage{mathtools}
\usepackage{url}
\usepackage{hyperref}
\usepackage{subfigure}
\usepackage{hhline}
\usepackage{color}
\usepackage{array,multirow}
\usepackage{bm}
\usepackage{cancel}
\newcommand{\beqa}{\begin{eqnarray}}
\newcommand{\eeqa}{\end{eqnarray}}
\newcommand{\beq}{\begin{equation}}
\newcommand{\eeq}{\end{equation}}

\newcommand{\nn}{\nonumber}
\newcommand{\bmt}{\begin{pmatrix}}
\newcommand{\emt}{\end{pmatrix}}
\usepackage[toc,page]{appendix}
\usepackage{comment}
\newcommand{\be}{\begin{equation}}
\newcommand{\ee}{\end{equation}}
\newcommand{\bea}{\begin{eqnarray}}
\newcommand{\eea}{\end{eqnarray}}

\begin{document}

\title{Inspection of new physics in $\rm B_s^0 \to K^+K^-$ decay mode}
\author{Manas K. Mohapatra}
\email{manasmohapatra12@gmail.com}




         
\affiliation{Department of Physics, IIT Hyderabad,
              Kandi - 502285, India }                            

              

\begin{abstract}
We scrutinize  the penguin dominated  $ B_s \to K^+ K^-$ decay mode involving  $b \to s$ quark level transition in family non-universal $Z^ \prime$ and vector-like down quark  model. There is discrepancy in the standard model branching ratio value of this mode  with the  experimental results reported by Belle, CDF and LHCb Collaborations. Additionally, the measured values of CP-violating asymmetries $ C_{K^+K^-}$ and $ S_{K^+K^-}$  deviate from the  SM results. We constrain the new parameter space by using the existing experimental limits on leptonic $ B_s\to \ell \ell$ ($\ell = e, \mu, \tau$)  processes.  We then check the effects of new physics on the branching ratio and CP violating parameters of the $ B_s \to K^+ K^-$ process.
\end{abstract}
\maketitle

\section{Introduction}
The in-depth search for physics beyond the standard model (SM) plays an important role in the area of particle physics. It is known that the CP asymmetry, the symmetry violation of combination of charge conjugation (C) and parity (P), is the main source for matter-antimatter asymmetry that is observed in our present universe. In the sector of quarks, the Cabbibo--Kobayasi--Maskawa(CKM) matrix  indicates a message for an insight to the gate way of CP violation particularly in B and K meson decays in the SM. However, it is not sufficient to understand the observed baryon asymmetry. Recently, various experimental hunts are going on to probe the physics beyond the SM. In this regard, B meson system provides an important role to study prominent observables like branching ratio, and  CP violating parameters  such as direct and mixing induced CP asymmetry to probe new physics. \par
We would like to study the $b\to s$ penguin dominated  $B_s^0 \to K^+ K^-$ decay mode which appears to have discrepancies in standard model values of CP-averaged branching ratio and CP violating parameters with the corresponding observed values.
The theoretical result for the observables are given in the TABLE-\ref{The-values}. Additionally, the TABLE-\ref{Exp-values} shows the results from Belle, CDF, LHCb Collaborations along with world averages. Thus these discrepancies between observed and predicted values could lead to probe the physics beyond the SM. 

In addition to this, the leptonic decays of pseudo-scalar meson $B_s^0$ sector plays a vital role and enthusiastically makes more attention to explore the physics beyond the SM. In particular, we study $B_s^0 \to \mu^+\mu^-$ decay mode because of careful observation of decay constant of neutral $B_s^0$ meson from lattice. On the other side,  the study of $B_s^0\to \ell \ell$ (where $\ell =e, \tau$)  put a less mark on the board as they have upper bounds. The former one has branching ratio with an upper limit of $2.8\times 10^{-7}$ (90\% C.L) \cite{Aaltonen:2009vr}, reported by LHCb where as $ < 6.8\times 10^{-3}$ (95\% C.L) reported by CDF Collaboration \cite{Aaij:2017xqt} for later decay mode.  The SM values of   branching ratio of $B_s^0 \to \tau ^+ \tau ^-$ and $B_s^0 \to e^+e^-$ decays have $\mathcal {O}(10^{-7})$ and $   \mathcal{O}(10^{-14})$ respectively where as for $B_s^0 \to \mu ^+ \mu ^-$, it is of the order of $10^{-9}$ \cite{Bobeth:2013uxa}. Thus there is a possibility of contribution to both decays along with $B_s^0 \to \mu^+ \mu^-$ mode in the new physics scenario.
 
Inspired by these discrepancies of the $B_s^0 \to K^+K^-$ decay mode, we would like to investigate, in QCD factorisation approach, the NP effect on CP-averaged branching ratio as well as the CP violation parameters  arising due to $Z^\prime$ model where an extra $U(1)^\prime$ gause boson $Z^\prime$ takes part in the play.  Several studies \cite{Li:2015xna, Mohanta:2008ce, Chiang:2006we, Mawlong:2008mb} have been done in the scenario of FCNC mediated by  $ Z^\prime$ boson.
And the new physics impact due to vector-like down quark(VLDQ) model \cite{Barenboim:2000zz, Atwood:2003tg, Giri:2003jj, Mawlong:2008mb} where an extra $SU(2)_L$ singlet down type quark has been added to SM quark sector which include a CP and flavor violating FCNC mediated by Z boson at tree level. The new coupling $ Z^\prime-b-s (Z-b-s)$ associated with $Z^ \prime$ (VLDQ) model can be constrained by using the experimental limit for all leptonic modes, and using the allowed parameter space, we examine the new physics impact on $B_s^0 \to K^+K^-$ decay mode observables.

\begin{table}  
\caption{The theoretical predictions of CP-averaged branching ratio ($\mathcal{B}$)(in the units of $10^{-6}$), CP-violating asymmetries such as direct ($\rm C_{KK}$ in $\%$) and mixing-induced  CP aymmetry ($\rm S_{KK}$).} \label{The-values}
\begin{tabular}{|l|l|l|l|l|l|}
\hline
\hline
$B_s^0\to K^+K^-$  & QCDF \cite{
Cheng:2009mu}  &  PQCD \cite{Wang:2014mua, Ali:2007ff}  &  SCET \cite{Williamson:2006hb} \\
\hline

$\mathcal{BR}$ &  25.2 $\substack{+12.7+12.5 \\ -7.2-9.1}$& \makecell{$13.6$\\ $19.7 \substack{+6.6 \\ -5.7}$}   &  18.2 \\
\hline

$\mathcal{C}_{KK}$ & -7.7 $\substack{+1.6 +4.0 \\-1.2-5.1}$  & \makecell{-23.3 \\ -16.4} & -6 \\ 

\hline
$\mathcal{S}_{KK}$ & 0.22$\substack{+0.04+0.05 \\-0.05-0.03}$  & \makecell{28 \\ 20.6} & 19 \\ 
\hline

\end{tabular}
\end{table}

\begin{table}  
\caption{Measured values of branching ratio ($\mathcal{B}$)(in the units of $10^{-6}$), CP-violating asymmetries such as direct ($C_{KK}$) and mixing-induced ($S_{KK}$) reported by Belle, CDF, LHCb collaborations and world averages.} \label{Exp-values}
\begin{tabular}{|l|l|l|l|l|l|}
\hline
\hline
$\bar{B}_s^0\to K^+K^-$  & Belle \cite{Peng:2010ze}  &  CDF \cite{Aaltonen:2011qt}  &  LHCb \cite{Aaij:2012as, Aaij:2018tfw, Aaij:2013tna}  & HFLAV \cite{Amhis:2016xyh}  & PDG \cite{Tanabashi:2018oca}\\
\hline

$\mathcal{BR}$ &  $38 \substack{+10 \\ -09}\pm 7$  & $23.9\pm1.4\pm3.6$   &  $23.0 \pm 0.7 \pm 2.3$  & $24.8\pm1.7$ & $25.9\pm1.7$ \\
\hline

$\mathcal{C}_{KK}$ & --  & -- & \makecell{$0.14 \pm 0.11 \pm 0.03$\\  $0.20 \pm 0.06 \pm 0.02$} & -- & $0.14 \pm 0.11$\\ 

\hline
$\mathcal{S}_{KK}$ & --  & -- & \makecell{$0.30 \pm 0.12 \pm 0.04$\\  $0.18 \pm 0.06 \pm 0.02$} & -- & $0.30 \pm 0.13$\\ 
\hline

\end{tabular}
\end{table}

The layout of this paper is structured  as follows. In section II, we discuss the effective Hamiltonian responsible for the non-leptonic  $b \to s q\bar q$ processes. We have also presented the framework for $B_s \to K^+ K^-$ observables such as  branching ratio and CP-violating parameters  in the standard model.  We constrain the new parameter space of $Z^ \prime$ model from the branching ratios of leptonic $B_s$ modes in  section III, and address the footprint of this model on $B_s \to K^+ K^-$ process by  using the new couplings. 
In section IV, we draw an attention to the interactions of the VLDQ model and check the impact on the aforementioned observables for $B_s \to K^+K^-$ decay mode. In the end, our results are summarized in  section V. 
\section{$B_s \to K^+ K^-$ process in the standard model}
In the standard model, the penguin dominated $B_s \to K^+K^-$ decay mode receives contribution from quark level transition $b\to s$ where the weak effective Hamiltonian describing the decay $b \to sq\bar{q}$ is given as \cite{Buchalla:1995vs}
\bea
\mathcal{H}_{\rm eff}=\frac{G_F}{\sqrt{2}}\Bigg \{V_{ub}V_{us}^*\bigg[C_1(\mu) O_1^{\mathit{u}}(\mu)+C_2(\mu) O_2^{\mathit{u}}(\mu)\bigg] -V_{tb}V_{ts}^*\bigg[\sum_{\mathit{i}=3}^{10}C_{\mathit{i}}(\mu) O_{\mathit{i}}(\mu)\bigg]\Bigg\}+ \rm h.c
\eea
where $\rm G_F=1.16639\times 10^{-5}~\rm{GeV}^{-2}$ is the Fermi coupling constant, $V_{\alpha \beta}$ are the CKM matrix element ($\alpha, \beta =u,b,s,t$). Here $O_{1,2}$ are the current-current operators, $O_{3,..6}$ are QCD penguin operators, $O_{7,..10}$  are electroweak penguin operators, and  $C_{\mathit{i}}(\mu)(\mathit{i}=1,...10)$  are the corresponding  Wilson coefficients evaluated at renormalization scale $\mu=m_b$ scale. 

Using the framework of QCD factorization approach \cite{Beneke:2003zv}, the decay amplitude can be  written in the form as
\bea
\langle K^+K^-|O_{\mathit{i}}|B_s^0 \rangle= \langle K^+K^-|O_{\mathit{i}}|B_s^0 \rangle _{fact}\big[1+\sum r_n \alpha _s^n+\mathit{O}(\frac{\Lambda _{QCD}}{m_b})\big],
\eea
where $\rm \langle K^+K^-|O_{\mathit{i}}|B_s^0 \rangle _{fact}$ represents the hadronic factorized matrix element, the second and third terms in the square bracket are higher order corrections. $\alpha _s$ is the strong coupling constant, $\rm \Lambda _{QCD}$ = 0.225 GeV is the QCD scale.

In the heavy quark limit, the amplitude of this decay mode can be represented as \cite{Beneke:2003zv},
\bea\label{SM Amp}
\mathcal{A}_{B_s^0 \to K^+K^-} &=&A_{K\bar{K}}\big[\delta_{\mathit{p}u} \alpha_1 + \alpha _4^{\mathit{p}}+\alpha_{4,EW}^{\mathit{p}}+ \beta _3^{\mathit{p}}+  \beta _4^{\mathit{p}}-\frac{1}{2} \beta _{3, EW}^{\mathit{p}}-\frac{1}{2}\beta_{4,EW}^{\mathit{p}}\big] \nn\\
& +& B_{\bar{K}K}\big[\delta _{\mathit{pu}}b_1+b_4^p+b_{4, EW}^{\mathit{p}}\big],
\eea
where 
\bea
&&A_{K\bar{K}}=\mathit{i} \frac{G_F}{\sqrt{2}} m_B^2 F_0^{B_s\to K}(0)f_K\,,\nn \\
&& B_{\bar{K}K}=\mathit{i}\frac{G_F}{\sqrt{2}}f_{B_s} f_K f_K\,,
\eea
 which includes the form factor $F_0^{B_s\to K}(0)$ at zero recoil momentum, and decay constants $f_{B_s}$ and $f_K$. The expressions of coefficients $\alpha _{\mathit{i}}$ and $\beta _{\mathit{i}}(b_i)$ are given in detail in ref. \cite{Beneke:2003zv}, which include factorisable along with non factorisable contributions to the above decay amplitude.

The CP-averaged branching ratio can be obtained using the following formula,
\bea
\mathcal{BR}(B_s^0 \to K^+K^-)=\tau _{B_s}\frac{p_c}{8 \pi m_{B_s}^2} \Bigg[\frac{{|\mathcal{A}_{B_s^0 \to K^+K^-}|^2}+{|\mathcal{A}_{\bar{B}_s^0 \to K^+K^-}|^2}}{2}\Bigg],
\eea
where $\tau _{B_s}$ ($m_{B_s}$) are the life time (mass) of $B_s$ meson and the center of mass (c.o.m) momentum in the rest frame of $B$ meson is given as 
\bea
p_c=\sqrt{(m_{B_s}^2-(m_{K^+}+m_{K^-})^2)(m_{B_s}^2-(m_{K^+}-m_{K^-})^2)}\,.
\eea
 The time dependent CP asymmetry of $B_s^0$ meson decaying to final CP eigenstate $K^+K^-$ can be written as \cite{Hayakawa:2013dxa}
\bea
\mathcal{A}_{K^+K^-}(t)=C_{KK}  \cos(\Delta M_{B_s^0}t)+ S_{KK}  \sin(\Delta M_{B_s^0}t)\,,
\eea
where $C_{KK}=\frac{|\lambda|^2-1}{1+|\lambda|^2}$ and  $S_{KK}=2\frac{\rm Im(\lambda)}{1+|\lambda|^2}$ are the direct and the mixing- induced CP asymmetries, respectively.  The parameter $\lambda$ is given as
\bea
\lambda=\frac{q}{p} \frac{\mathcal{A}_{{\bar{B}_s^0 \to K^+K^-}}}{\mathcal{A}_{{B_s^0 \to K^+K^-}}},
\eea
where, $q$ and $p$ are mixing parameters which are connected to the standard model CKM elements as
\bea
\frac{q}{p}=\frac{V_{tb}^*V_{ts}}{V_{tb}V_{ts}^*}\,.
\eea
Symbolically, the amplitude of the $B_s\to K^+K^-$  decay mode can be written as
\bea \label{SM-amp(parameterized)}
\mathcal{A}_{{B_s^0 \to K^+K^-}} & =\zeta _\mathit{u} \mathcal{A}_\mathit{u}+ \zeta _\mathit{c} \mathcal{A}_\mathit{c} \nn \\
& = \zeta _\mathit{c} \mathcal{A}_\mathit{c}\big[1+\wp a e^{\mathit{i}(\delta _1-\gamma)}\big]
\eea
where $\zeta _\mathit{q}=V_{qb}V_{qs}^*$($q=u, c$),  $a=|\frac{\zeta _\mathit{u}}{\zeta _\mathit{c}}|$, $ \wp=|\frac{\mathcal{A}_\mathit{u}}{\mathcal{A}_\mathit{c}}|$
, $\gamma$ is the weak phase of CKM element $V_{ub}$, and $\delta_1$ is the relative weak phase between $\mathcal{A}_\mathit{u}$ and $\mathcal{A}_\mathit{c}$.
From the amplitude given in \ref{SM-amp(parameterized)}, the parameters $\mathcal{BR}$, $\mathcal{ C}_{KK}$ and $\mathcal {S}_{KK}$ can be obtained respectively as
\begin{align}
\mathcal{BR} &=\frac{\tau_{B_s}p_c}{8\pi m_{B_s}^2}|\zeta _\mathit{c} \mathcal{A}_\mathit{c}|^2\big\{1+(\wp  a)^2+2\wp a \cos \delta_1  \cos \gamma\big\}, \\
\mathcal{C}_{KK}&=-\frac{2\wp a \sin \delta_1 \sin \gamma}{1+(\wp a)^2+2\wp a \cos \delta_1 \cos \gamma}, \\
\mathcal{S}_{KK}&=\frac{\sin 2 \beta +2\wp a \cos \delta_1 \sin (2\beta - \gamma)+(\wp a)^2 \sin (2\beta-2\gamma)}{1+(\wp a)^2+2\wp a \cos \delta_1 \cos \gamma}.
\end{align}
For numerical computation of these observables in the SM, the CKM matrix elements along with the weak angle $\gamma$, all particle masses and the life time of $B_s$ meson are taken from \citep{Tanabashi:2018oca}. We use the value of $\beta _s$  angle from \citep{Koppenburg:2017mad}. The calculated Wilson coefficients in naive dimensional regularization scheme at NLO are taken from \citep{Buchalla:1995vs} at $m_b$ scale. The value of form factor $F_{B_s}(0)$ at zero recoil momentum and the decay constant $f_{B_s}$ are taken from \cite{Khodjamirian:2017fxg}. In addition to this the decay constant  $f_K$ is taken from \cite{Lu:2018cfc}. Using these input values, 
the predicted SM values of the $B_s \to K^+ K^-$ observables are given as
\bea
&&\mathcal{BR}=(34.37 \substack {+7.90\\-5.61})\times 10^{-6},\,\\
&&C_{KK}=-0.11 \substack{+0.0168\\-0.0151},\,\\
&&S_{KK}=0.32 \substack {+0.042\\-0.035}\,.
\eea
Here, the theoretical uncertainties for the above observables mainly arise  from form factor, decay constants \cite{Khodjamirian:2017fxg, Lu:2018cfc},  CKM matrix elements \citep{Tanabashi:2018oca}.
\section{$Z^ \prime$ model}
In this section we discuss the effects of new physics associated with $Z^\prime$ model on the observables of $B_s \to K^+ K^-$ decay process. We constrain the $Z^\prime$ new couplings by using the experimental limits on  $B_s \to \ell \ell$  (where $\ell$ is any charged leptons), mediated by the FCNC transitions $b \to s \ell \ell$. These are the theoretically cleanest $B$ decays as  the only non-perturbative quantity involved in the description of these processes is the $B_s$ meson decay constant. 
\subsection{$B_s \to \ell ^+\ell^-(\ell =e, \mu , \tau)$ processes} 
\begin{figure}[htb]
\centering
\includegraphics[scale=0.7]{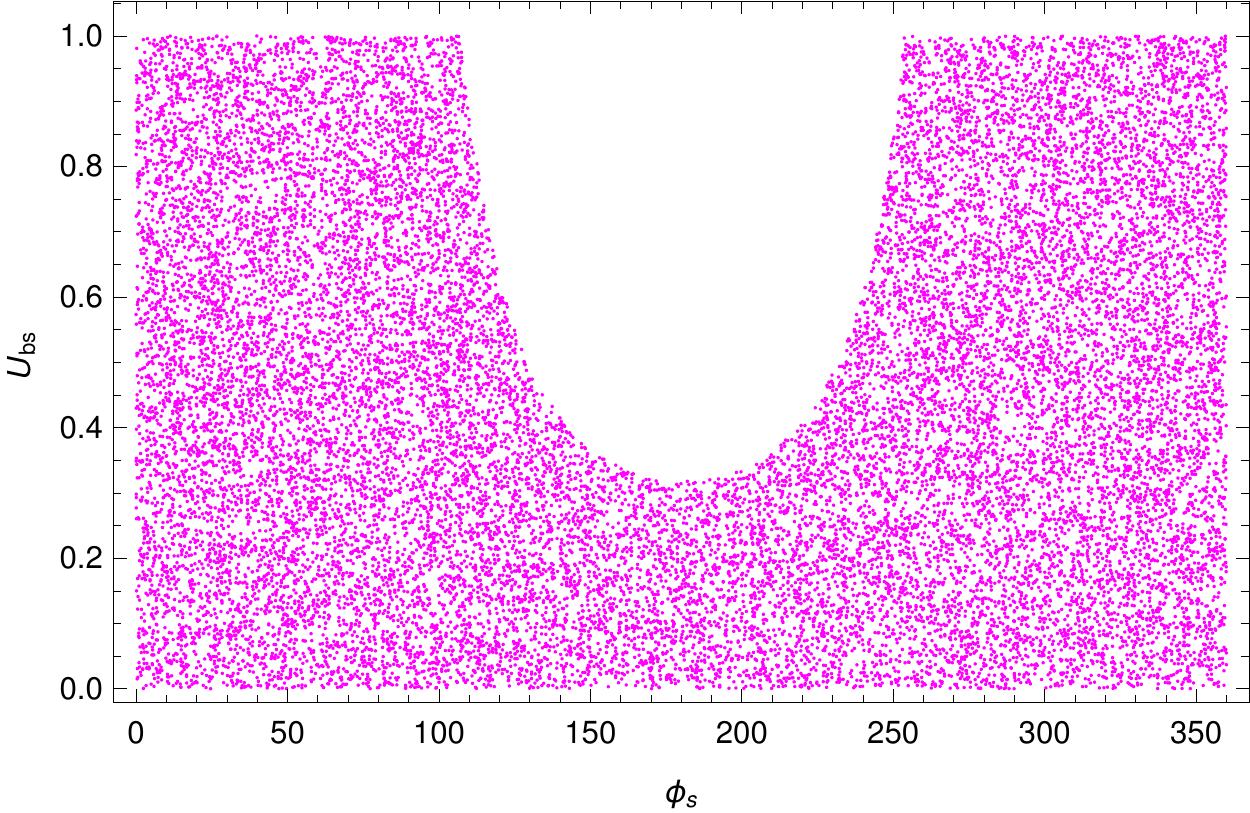} 
\caption{Constraints on new coupling parameter from the branching ratios of leptonic $B_s\to \ell ^+ \ell ^-(\ell =e, \mu, \tau)$ processes in $Z^\prime$ model.}\label{Fig:result1}
\end{figure}

In the SM, the effective Hamiltonian for quark level transitions $b \to s \ell ^+ \ell ^-$ is given by \cite{Beneke:2004dp, Sahoo:2016edx}
\bea\label{Heff leptonic}
\mathcal{H}_{\rm {eff}}=-\frac{G_F}{\sqrt{2}}\big[\lambda_t^{(q)}\mathcal{H}_{\rm {eff}}^{(t)} + \lambda_u^{(q)}\mathcal{H}_{\rm {eff}}^{(u)} \big] + h.c.,
\eea
where
\bea\label{inf}
\mathcal{H}_{\rm {eff}}^{(u)}&=&C_1(\mathcal{O}_1^c-\mathcal{O}_1^u) + C_2(\mathcal{O}_2^c-\mathcal{O}_2^u), \nn\\
\mathcal{H}_{\rm {eff}}^{(t)}&=&C_1\mathcal{O}_1^c+C_2 \mathcal{O}_1^u) + \sum_{\mathit{i}=3}^{10}C_i\mathcal{O}_i,
\eea
$\lambda _k^{(q)}=V_{kb} V_{kq}^*$ and $C_i$'s are the Wilson coefficients.
Using effective Hamiltonian \ref{Heff leptonic}\,, the transition amplitude  for this process is given as 
\bea \label{SM amp}
\mathcal{M}(B_s\to \ell^+ \ell^-)=\mathit{i} \frac{G_F\alpha}{\sqrt{2}\pi} V_{tb}V_{ts}^*f_{B_s} C_{10}m_\ell (\bar{\ell}\gamma _5 \ell), 
\eea
where $\alpha$ is the fine structure constant. Here, we have used the vacuum insertion method to define the decay constant in the matrix element as
\bea 
<0|\bar{s}\gamma^ \mu \gamma _5 b|B_s^0>=if_{B_s}p_B^\mu,  
\eea
where $p_B^{\mu} =p_{\ell+}^\mu+p_{\ell-}^\mu$. In general, from equation \ref{SM amp}\,, the associated branching ratio is given as \cite{Mohanta:2005gm}
\bea\label{Br SM}
\mathcal{BR}(B_s\to \ell^+\ell^-)=\frac{G_F^2\tau_{B_s}}{16 \pi ^3}\alpha^2f_{B_s}^2 m_{B_s} m_\ell ^2|V_{tb}V_{ts}^*|^2| C_{10}|^2
\sqrt{1-\frac{4m_\ell^2}{m_{B_s}^2}}
\eea
Using the $B_s$ decay constant from \cite{Khodjamirian:2017fxg} and remaining input parameters from PDG \cite{Tanabashi:2018oca}, the predicted SM branching ratio values are presented below. The errors in the SM results are coming mainly from decay constants and  CKM matrix elements. Here we also show  the corresponding experimental limits for all leptonic decay modes  \cite{Tanabashi:2018oca}\,.
\bea\label{SM br}
\mathcal{BR}_{B_s\to \mu^+\mu^-}^{SM}&=&(3.47\substack{+1.21\\-0.85})\times 10^{-9},
\hspace{1cm}\mathcal{BR}_{B_s\to \mu^+\mu^-}^{Exp}= (2.7 \substack{+0.6\\ -0.5}) \times 10^{-9}\,, \nn\\
\mathcal{BR}_{B_s\to \tau^+\tau^-}^{SM}&=&(7.44\substack{+2.6\\-1.82})\times 10^{-7},
\hspace{1cm} \mathcal{BR}_{B_s\to \tau^+\tau^-}^{Exp}< 6.8 \times 10^{-3}\,,
\nn\\
\mathcal{BR}_{B_s\to e^+e^-}^{SM}&=&(8.19\substack{+2.86\\-2.01})\times 10^{-14},
\hspace{1cm} \mathcal{BR}_{B_s\to e^+e^-}^{Exp}< 2.8 \times 10^{-7}\,.
\eea
Though $B_s\to \ell^+\ell^-$ decays occur only at one-loop level in the SM, these processes can occur at tree level in the presence of new $Z^\prime$ gauge boson arising due to the $U(1)^\prime$ gauge extension of the SM. 
The effective Hamiltonian corresponding to the transition $b \to s \ell ^+ \ell^-$  process is given by \cite{Chang:2009tx, Barger:2009qs}
\bea \label{Bs-Br-ZP}
\mathcal{H}_{\rm eff}^{Z^\prime} &=& -\frac{2G_F}{\sqrt{2}}V_{tb}V_{ts}^* \Big (\frac{g^ \prime M_Z}{g_1 M_Z^\prime} \Big)^2 \bigg[\frac{U_{bs}^LU _{\ell \ell}}{V_{tb}V_{ts}^*}(\bar{s}b)_{V-A}(\bar{\ell} \ell)_{V-A}\bigg] \nn\\
&-& \frac{U_{bs}^L U_{\ell \ell}^R}{V_{tb}V_{ts}^*}(\bar{s}b)_{V-A}(\bar{\ell} \ell)_{V+A} + h.c.,
\eea
where $g_1 (g^\prime)$ is the coupling constant of $Z$($Z^\prime$) boson. According to the SM effective Hamiltonian \ref{Heff leptonic}\,, the Hamiltonian in $Z^\prime$ can be written as 
\bea
\mathcal{H}_{\rm eff}^{Z^\prime}=-\frac{G_F}{\sqrt{2}}V_{tb}V_{ts}^*\bigg[C_9^{Z^\prime}O_9 + C_{10}^{Z^\prime}O_{10}\bigg] + \rm h.c,
\eea
where the new Wilson coefficients are given as
\bea
C_9^{Z^\prime}&=&-2\Big( \frac{g^ \prime M_Z}{g_1 M_{Z^\prime}}\Big)^2 \frac{U_{bs}^L}{V_{tb}V_{ts}^*}(U_{\ell \ell}^L+U_{\ell \ell}^R)=-2\Big( \frac{g^ \prime M_Z}{g_1 M_{Z^\prime}}\Big)^2 \frac{|U_{bs}^L| e^{i\phi_s}}{V_{tb}V_{ts}^*}(U_{\ell \ell}^L+U_{\ell \ell}^R), \nn\\
C_{10}^{Z^\prime}&=&2\Big( \frac{g^ \prime M_Z}{g_1 M_{Z^\prime}}\Big)^2 \frac{U_{bs}^L}{V_{tb}V_{ts}^*}(U_{\ell \ell}^L-U_{\ell \ell}^R)=2\Big( \frac{g^ \prime M_Z}{g_1 M_{Z^\prime}}\Big)^2 \frac{|U_{bs}^L| e^{i\phi_s}}{V_{tb}V_{ts}^*}(U_{\ell \ell}^L-U_{\ell \ell}^R)\,,
\eea
with  $\phi _s$ is the associated weak phase of $U_{bs}$. 
We consider $\frac{g^\prime}{g_1} \sim 1$ with the assumption that both the $U(1)$ groups have same origin from some grand unified theory. For a TeV-scale $Z^\prime$ boson, their ratio of masses $M_Z/M_{Z^\prime}$ will be $\sim 10^{-1}$. In this analysis, the coupling of $Z^\prime$ boson to leptons $U_{\ell \ell}^{L(R)}$ are considered to be SM-like.
Now, comparing the theoretical values of $B_s\to \ell \ell$ branching ratios with their corresponding $1\sigma$ range of experimental data, we constrain the new  $Z^\prime-b-s$ coupling ($U_{bs}$) and weak phase $(\phi_s)$ as shown in FIG. \ref{Fig:result1}\,. From this figure, the allowed range of $U_{bs}$ and $\phi_s$ parameters of $Z^\prime$ model  are given as   
\bea \label{ZP-Con}
1.2\times 10^{-5} \leq |U_{bs}|\leq 0.99,  \hspace{0.5cm}  0^ {\circ}\leq \phi _s \leq 360^ {\circ}\,.
\eea

\subsection{Impact on non-leptonic $B_s \to KK$ decay mode}
Now, we will  discuss the impact of family non-universal $Z^\prime$  gauge boson on the $B_s \to K^+K^-$ decay mode.
The effective Hamiltonian describing the $b \to s q \bar{q}$ (q=u,d) transition for $B_s\to K^+K^-$ decay mode is given as \cite{Mohanta:2008ce} 
\bea\label{Z' eff}
\mathcal{H}_{\rm eff}^{Z^\prime}= \frac{2G_F}{\sqrt{2}}\bigg(\frac{g'M_Z}{g_1M_{Z'}}\bigg)^2U_{bs}^L(\bar{s}b)_{V-A}\sum _q\big[U_{qq}^L(\bar{q}q)_{V-A}+U_{qq}^R(\bar{q}q)_{V+A}\big]+h.c..
\eea 
 Due to hemiticity of effective Hamiltonian, the diagonal elements of $U_{qq}^{L(R)}$ are real where as the non-diagonal elements of $U_{bs}^{L(R)}$ might be complex along with a phase $\phi _s$.
Now, comparing the effective Hamiltonian of $Z^\prime$ model \ref{Z' eff} with the general effective Hamiltonian   
\bea
\mathcal{H}_{\rm eff}=-\frac{G_F}{\sqrt{2}}V_{tb}V_{ts}^*\sum _q\big[C_9^\prime O_9+C_7^\prime O_7\big]\,,
\eea 
we obtain
\bea
C^\prime _{9(7)}=-\frac{4}{3}\bigg(\frac{g'M_Z}{g_1M_{Z'}}\bigg)^2U_{bs}^L \bigg(U_{uu}^{L(R)}-U_{dd}^{L(R)}\bigg).
\eea
where $C_9^\prime, C_7^\prime$ are the new Wilson coefficients arising due to $Z^\prime$ gauge boson. 
Many studies have been done in \cite{Langacker:2000ju, He:2006bk, Baek:2008vr, Mohanta:2008ce, Everett:2009cn, Barger:2004qc} with the manifestation of electroweak contribution assuming $U_{uu}^{L(R)} \simeq  -2U_{dd}^{L(R)}$. Thus, 
\bea
C^\prime _{9(7)}=4\bigg(\frac{g'M_Z}{g_1M_{Z'}}\bigg)^2\frac{U_{bs}^L U_{dd}^{L(R)}}{V_{tb}V_{ts}^*}.
\eea
Now for convenience, these coefficients can be written in the following parametric form as
\bea
C_{9(7)}^\prime =4\bigg(\frac{g'M_Z}{g_1M_{Z^ \prime}}\bigg)^2U_{bs}^L=4\bigg(\frac{g'M_Z}{g_1M_{Z^\prime}}\bigg)^2\frac{|U_{bs}^L|e^{\mathit{i}\phi _s}}{V_{tb}V_{ts}^*}\,,
\eea
where the assumption of $U_{qq}^{L(R)}\sim 1$ has been taken out from experimental data of $B_s$ meson \cite{Li:2015xna}. The decay amplitude in presence of additional $Z^\prime$ boson can be written as
\bea\label{NP Amp Z'}
\mathcal{A}_{B_s^0 \to K^+K^-} &=&A_{K\bar{K}}\big[\delta_{\mathit{p}u} \alpha_1 + \alpha _4^{\mathit{p}}+\alpha_{4,EW}^{\mathit{p}}+ \beta _3^{\mathit{p}}+  \beta _4^{\mathit{p}}-\frac{1}{2} \beta _{3, EW}^{\mathit{p}}-\frac{1}{2}\beta_{4,EW}^{\mathit{p}}\big] \nn\\
& +& B_{\bar{K}K}\big[\delta _{\mathit{pu}}b_1+b_4^p+b_{4, EW}^{\mathit{p}}\big] \nn\\
&-&\zeta _t \Big [ A_{K\bar{K}}\big (\tilde{ \alpha} _4^{\mathit{p}}+\tilde{\alpha}_{4,EW}^{\mathit{p}}+ \tilde{\beta} _3^{\mathit{p}}-\frac{1}{2} \tilde{\beta} _{3, EW}^{\mathit{p}} \big) \Big ]\,,
\eea
where terms $\tilde{\alpha}$ and $\tilde{\beta}$  arise due to new physics contributions.
We can represent the above transition amplitude in the parametrized form as
\bea
 \mathcal{A}=A^{SM}-\zeta _t A^{NP}=\zeta _c A_c\big[1+\wp ae^{i(\delta _1-\gamma)}-\wp ^\prime b e^{(\delta _2 +\phi _s)}\big]\,.
\eea
In addition to $\wp$ and $a$, given in the previous section,  the new parameters in the above amplitude are defined as
$b=|\frac{\zeta _\mathit{t}}{\zeta _\mathit{c}}|$, $ \wp^\prime=|\frac{\mathcal{A}_\mathit{NP}}{\mathcal{A}_\mathit{c}}|$, and $\delta_2$ is the relative strong phase. 

The CP-averaged branching ratio can be written as
\bea\label{NP BR Z'}
\langle \mathcal{BR}\rangle &=&\frac{\tau _{B_s} p_c}{8 \pi m_{B_s}^2} |\zeta _c A_c|^2\bigg[\mathcal{G}+2\wp a\cos \delta _1 \cos \gamma -2 \wp ^\prime b\cos \delta _2 \cos \phi _s\nn \\&&-2 \wp \wp ^\prime ab \cos(\delta _1 -\delta _2) \cos (\gamma + \phi _s)\bigg] \,.    
\eea
The direct CP asymmetry is given as
\bea
\mathcal{C}_{KK}=
-\frac{2\big[ra \sin \delta _1 \sin \gamma + \wp ^\prime b \sin \delta _2 \sin \phi _s +\wp  \wp ^\prime a b \sin (\delta _2-\delta _1)\sin( \gamma +\phi _s)\big]}{\big[ \mathcal{G}+2\big(\wp a \cos \delta _1 \cos \gamma - 2 \wp ^\prime b\cos \phi _s \delta _2-2 \wp \wp ^\prime b b ^\prime\cos (\gamma + \phi _s)\cos (\delta _2 -\delta _1)\big)\big]}\,,
\eea
The mixing induced CP asymmetry can be represented as
\bea
\mathcal{S}_{KK}=\frac{\mathcal{M}}{\mathcal{G}+2 \wp a \cos \delta _1 \cos \gamma -2\wp ^\prime b \cos \delta _2 \cos \phi _s-2 \wp \wp ^\prime a b \cos (\delta_1 -\delta _2) \cos(\gamma + \phi _s)},
\eea
where $\mathcal{G}=1+(\wp a)^2+ (\wp ^\prime b)^2$ 
and
\begin{align}
\mathcal{M}=&\sin 2 \beta +2 \wp a \cos \delta _1 \sin(2 \beta +\gamma)-2 \wp ^\prime  b \cos \delta _2 \sin(2 \beta -\phi _s)+(\wp a)^2\sin (2 \beta +2 \gamma) \\ 
+& (\wp ^\prime b)^2\sin(2 \beta -2 \phi _s)-2 \wp \wp ^\prime a b \cos(\delta -\delta ^\prime) \sin(2 \beta + \gamma -\phi _s).
\end{align}



\begin{figure}[htb]
\centering
\includegraphics[scale=0.85]{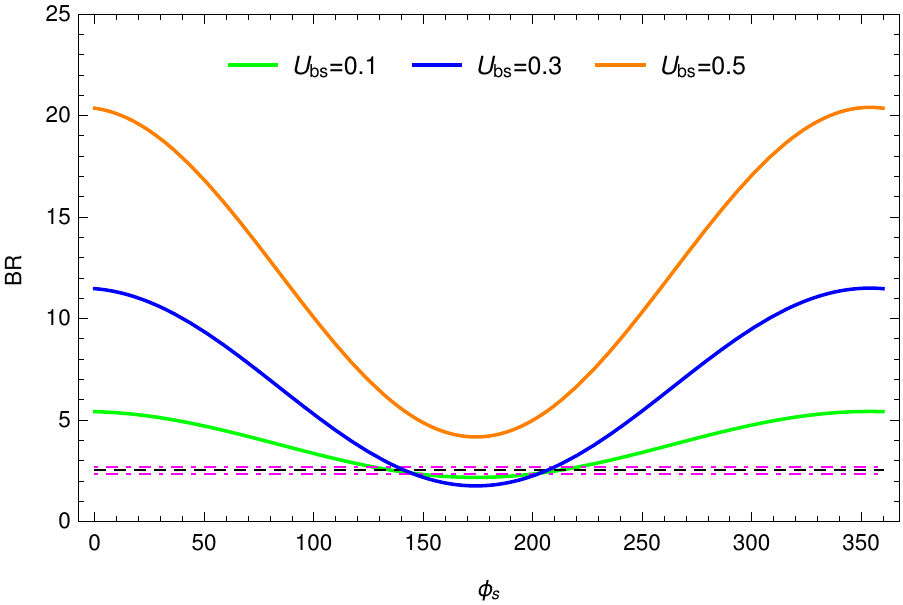} 
\quad
\includegraphics[scale=0.85]{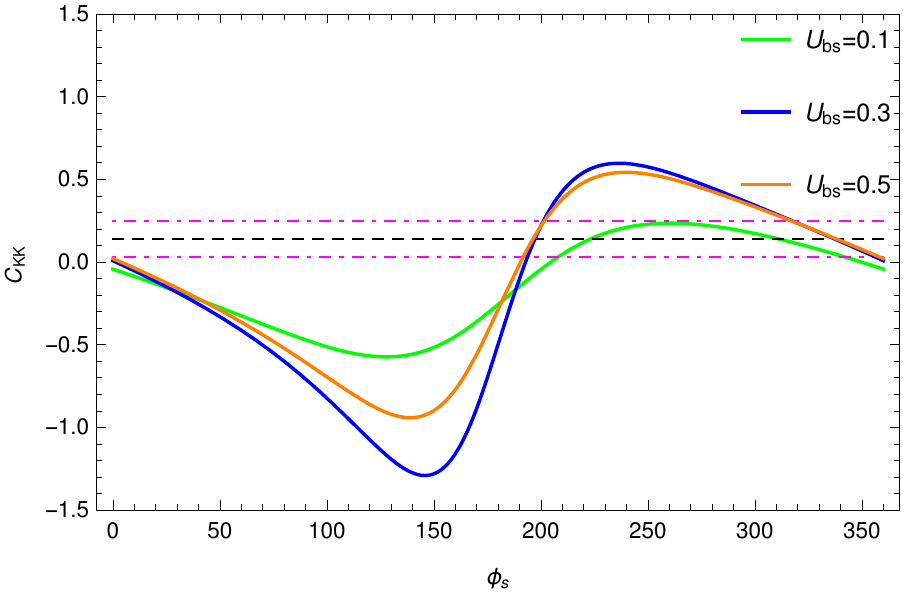} 
\quad
\includegraphics[scale=0.85]{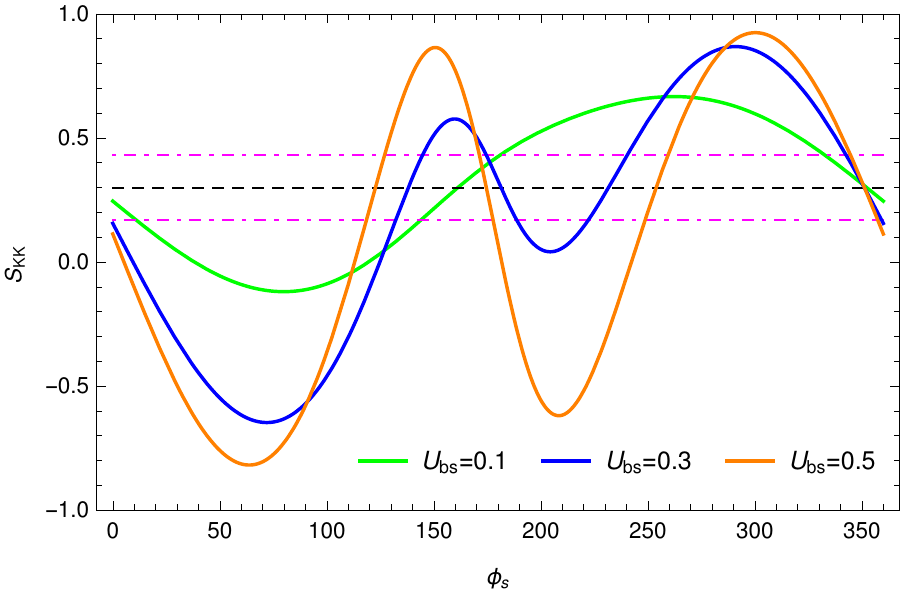} 
\caption{Variation of CP-averaged branching ratio (in the units of $10^{-5})$ (top-left panel), direct CP asymmetry (top-right panel) and mixing induced CP asymmetry (bottom panel) (in \%) with the new weak phase $\phi _s$ for different $|U_{bs}|$ entries. The black color horizontal dotted line represents the central experimental value where as the magenta dotted lines denote the 1$\sigma$ error limit.}
\label{Fig:result2}
\end{figure}

\begin{figure}[htb]
\centering
\includegraphics[scale=0.6]{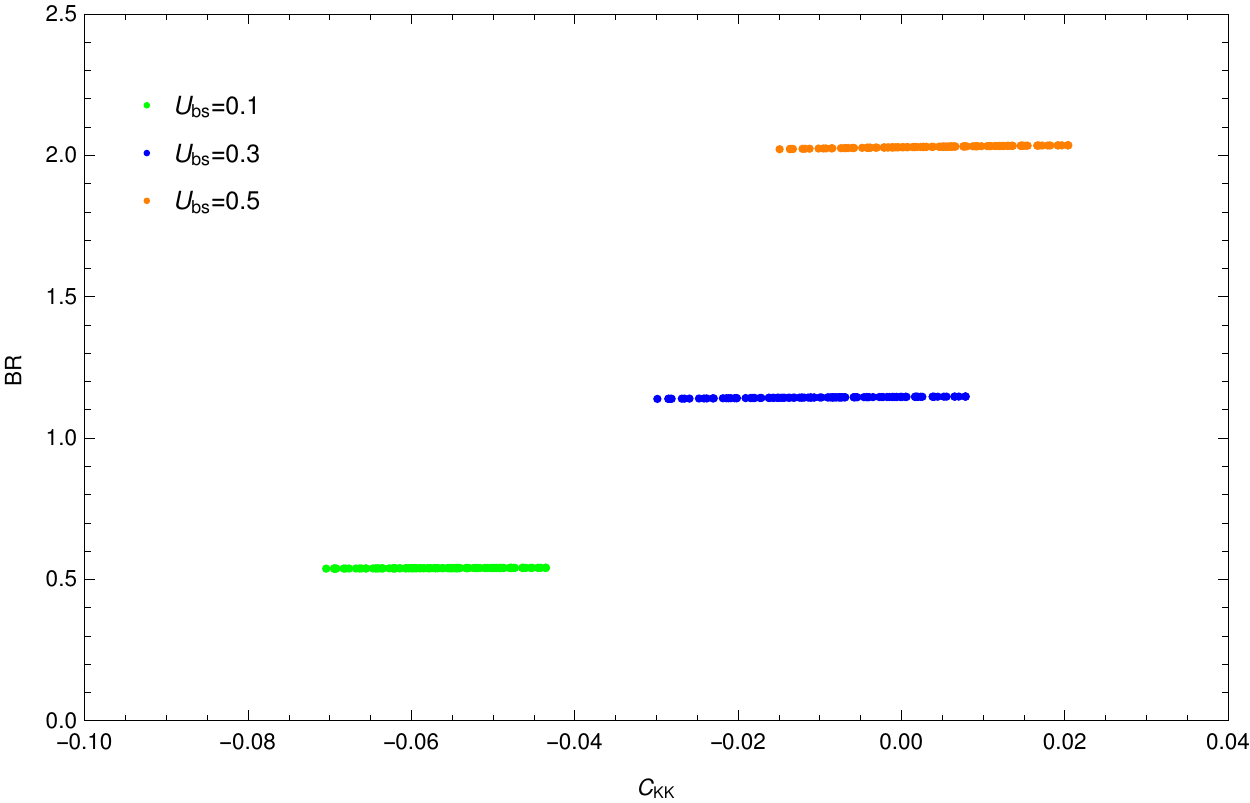} 
\quad
\includegraphics[scale=0.6]{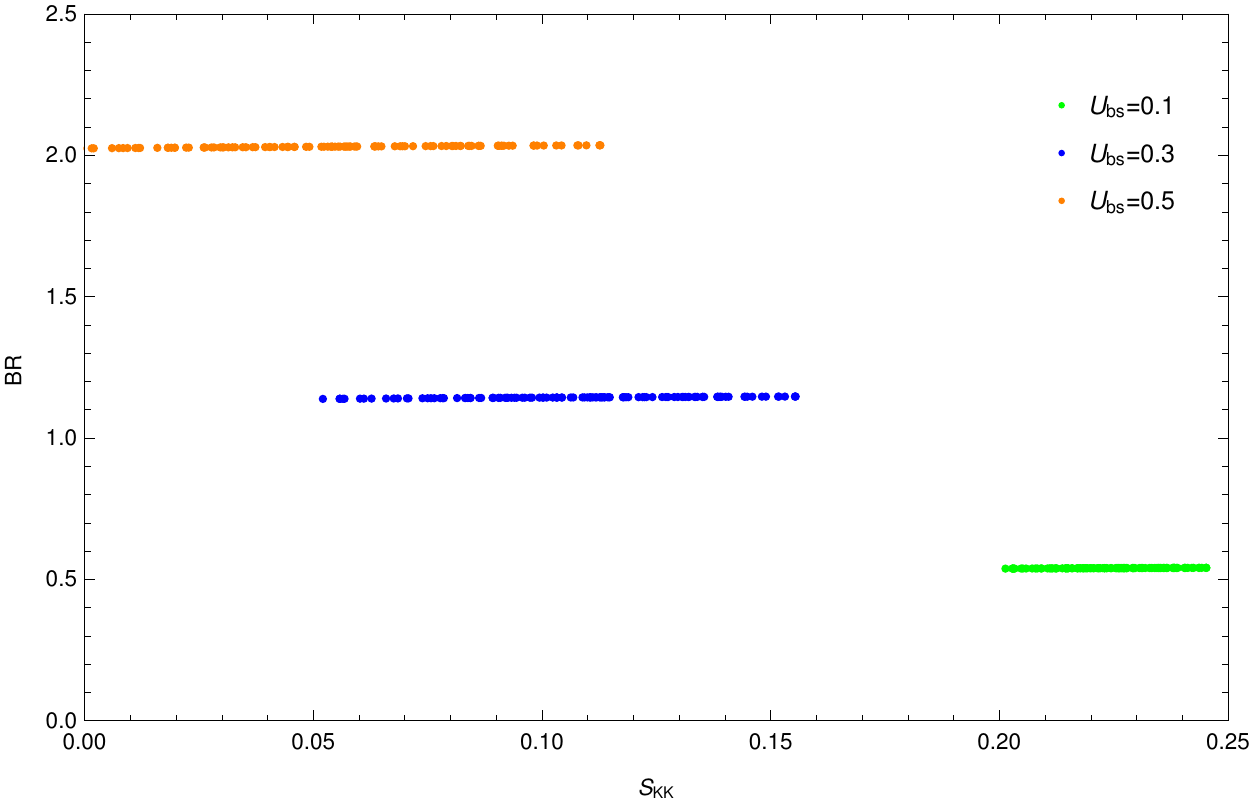} 
\quad
\includegraphics[scale=0.6]{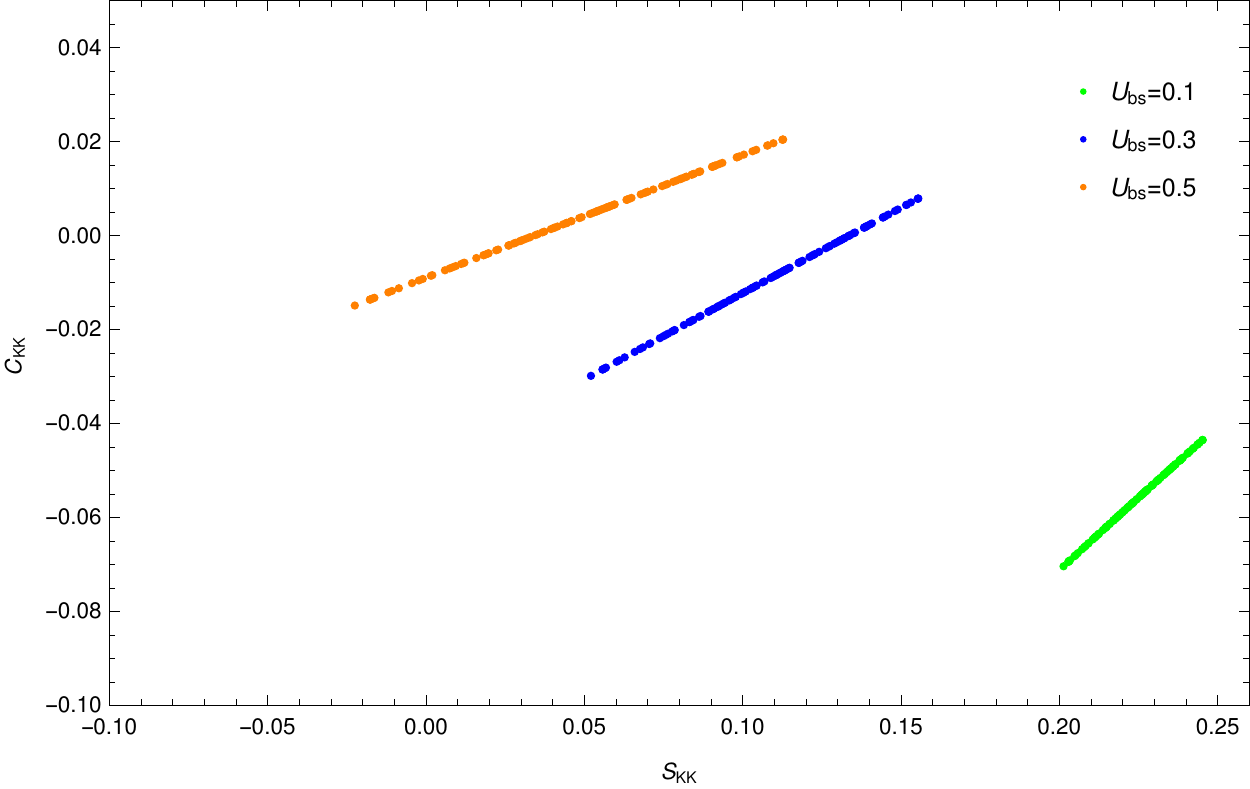} 
\caption{Correlation plots of CP-averaged branching ratio (in the units of $10^{-4})$ with direct CP asymmetry (top-left panel) and mixing induced CP asymmetry (top-right panel) (in \%), and direct CP asymmetry with mixing induced CP asymmetry (bottom panel)}
\label{Fig:result3}
\end{figure}

After collecting the detailed expression for CP-averaged branching ratio and the CP violating parameters in the presence of new $Z^\prime$ gauge boson, we now proceed for numerical analysis for these observables. Using the allowed  parameter space  from equation \ref{ZP-Con}\,  we show the variation of CP-averaged branching ratio (top-left panel), direct CP violation (top-right panel) and mixing induced CP asymmetry (bottom panel) with respect to mixing weak phase $\phi _s$ with some benchmark entries of $\rm U_{bs}$(color) as 0.1(Green), 0.3(Blue) and 0.5(Orange) in FIG. \ref{Fig:result2}\,.  As we see from the top-left one,  for both blue and green colored $\rm U_{bs}$ values, the NP effect of branching ratios lie in $1\sigma$ range where as the one with $0.5$ value,  is not accommodating in the same error limit. On the other side, if we observe the CP violating parameters in the new physics scenario,  both $\rm C_{KK}$ and $\rm S_{KK}$ lie within $1 \sigma$ experimental data.
 FIG. \ref{Fig:result3} depicts the  correlation among all the discussed observables. In this figure, the top-left (right) panel displays $\rm BR-\rm C_{KK}~(\rm S_{KK})$ correlations and  the plot in the bottom panel shows a relationship between $\rm C_{KK}$ and $\rm S_{KK}$. We have shown the predicted results of branching ratios, CP violating observables  for different values of $\ U_{bs}$ and $\phi _s$ in the top-section of Table \ref{Tab:result}\,.
\section{Vector like down quark (VLDQ) model}
Here we study the minimal extension of SM where the quark sector is expanded by an extra vector-like down quark. Because of this, we obtain a $4 \times 4$ matrix $V_{\mathit{i}\alpha}$ ( $\mathit{i}=\mathit{u, c, t, t ^\prime}$  and $\alpha =d, s, b, b ^\prime$) from which the interaction of this extra down-type quark with the SM quarks could be obtainable and scrutinize the deviations of the unitarity  relation of the CKM matrix. This mixing provides a remarkable study of flavor changing neutral current (FCNC) interaction where $Z$ particle is mediated  through tree level contribution.
 In general, this model include the following Lagrangian \cite{Giri:2003jj}.
 $$\mathcal{L}_Z=\frac{g}{2 \cos \theta _W}\big[\bar{\mathit{u}}_{L\mathit{i}} \gamma ^ \mu \mathit{u}_{L\mathit{i}} -\bar{d} _{L \alpha} Q_{\alpha \beta} \gamma ^\mu d_{L \alpha} -2 \sin^2 \theta_W  J_{em}^\mu\big] Z_ \mu$$
 where  $L$ denote the left handed chiral particles, $i$  and $\alpha, \beta$  denote the generation indices for up-type and down-type quarks respectively.  
The second term in the above Lagrangian corresponds to the mixing in the down-type quark sector and the matrix $Q_{\alpha \beta}$  can be represented as 
\bea\label{Unitarity}
Q_{\alpha \beta}=\sum_{\mathit{i}=\mathit{u, c, t}}V_{\alpha \mathit{i}}^\dagger V_{\mathit{i}\beta}=\delta_{\alpha \beta}-V_{4 \alpha}^*V_{4 \beta}
\eea
Here $V$ is not unitary as an extra down-type vector like quark of charge ($-\frac{1}{3}$) has been added to the SM. 
It provides a new signal to probe the physics beyond the SM and modify the CP asymmetries and branching ratio predictions. We constrain the  new parameters from the Br($B_s \to \ell^+ \ell^-$), to be presented
in the subsequent section.
\subsection{$B_s \to \ell^+ \ell^-(\ell =e, \mu, \tau)$ processes}
\begin{figure}[htb]
\centering
\includegraphics[scale=0.7]{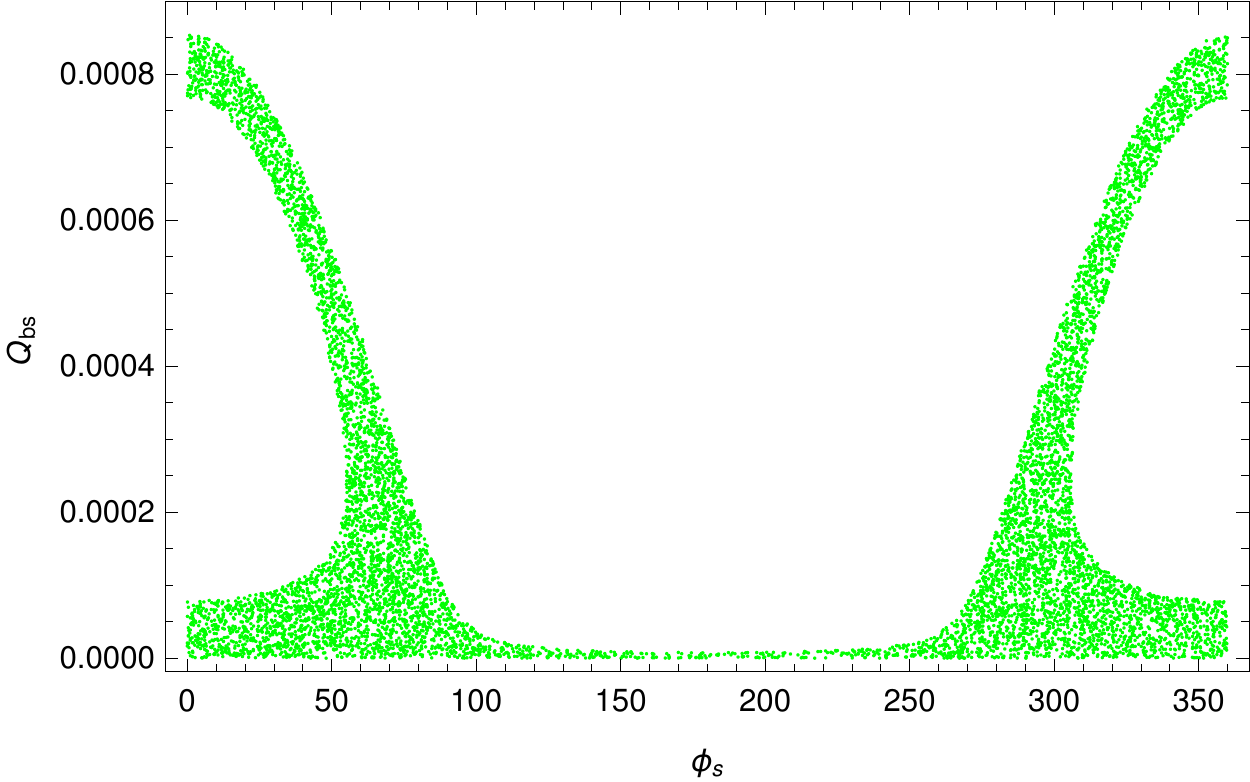} 
\caption{Constraints on new coupling parameter from all the branching ratios of leptonic $B_s\to \ell ^+ \ell ^-(\ell =e, \mu, \tau)$ processes in VLDQ model.}\label{Fig:result4}
\end{figure}

Though $B_s\to \ell^+\ell^-$ process are suppressed in the SM, but can be significant  in the presence of extra vector like down quark particle where $Z$ is mediated at tree level whose contribution provides the physics beyond the SM.  
The branching ratio of $B_s\to \ell^+\ell^-$ in $Z$ mediated VLDQ model is given by \cite{Chen:2010aq}
\bea\label{Br VLDQ}
\mathcal{BR}(B_s\to \ell^+\ell^-)=\frac{G_F^2\alpha ^2m_{B_s}m_\ell ^2f_{B_s}^2 \tau_{B_s}}{16\pi ^3} |V_{tb}V_{ts}^*|^2\sqrt{1-4(\frac{m_\ell ^2}{m_{B_s}^2})}\left|C_{10}^{\rm tot}\right |^2,
\eea
where 
\bea
C_{10}^{\rm tot}=C_{10}-\frac{\pi}{\alpha}\frac{Q_{bs}}{V_{tb}V_{ts}^*}\,.
\eea
Here the second term defines the new parameter when b quark couples to s quark at tree level.
Using the theoretical and experimental values of  $B_s \to \ell \ell$ \ref{SM br}\,, the constraint on $Q_{bs} -\phi _s$ parameters are presented in FIG. \ref{Fig:result4}\,. The  ranges obtained from the constrained plot are given as
\bea
0\leq |Q_{bs}|\leq 8.52\times 10^{-4}, \hspace{0.5cm} 0^ {\circ}\leq \phi _s \leq 360^ {\circ}.
\eea

\subsection{Effect on $B_s \to KK$ mode}
The effective Hamiltonian corresponding to new interaction describing $b\to sq \bar{q}$ can be represented as
$$\mathcal{H}_{eff}^Z=-\frac{G_F}{\sqrt{2}}V_{tb}V_{ts}^*\big[\tilde{C}_3O_3+\tilde{C}_7O_7+\tilde{C}_9O_9\big],$$
where the new Wilson coefficients at the $M_Z$ scale are given as \cite{Atwood:2003tg, Deshpande:2003nx}
\bea \label{new couplings}
\tilde{C}_3(M_Z)&=&\frac{1}{6}\frac{Q_{bs}}{V_{tb} V_{ts}^*},\nn\\
\tilde{C}_7(M_Z)&=&\frac{2}{3}\frac{Q_{bs}}{V_{tb} V_{ts}^*}\sin^2\theta _W\,, \nn\\
\tilde{C}_9(M_Z)&=&-\frac{2}{3}\frac{Q_{bs}}{V_{tb} V_{ts}^*}(1-\sin^2\theta _W)\,.
\eea
Here  $Q_{bs}=|Q_{bs}|e^{\mathit{i} \phi_s}$ and $\sin^2 \theta _W=0.231$. As the new couplings are in $M_Z$ scale, so these can be evolved to $m_b$ scale employing renormalization group equation \cite{Buchalla:1995vs}. 
By using RGE, these three couplings can be generated and we consider the above values from  the ref.  \cite{Mohanta:2008fa}. 

Now using the unitarity condition from equation \ref{Unitarity}\,, we get 
\bea
\xi _u +\xi _c + \xi _t =Q_{bs}\,.
\eea
To this relation,  we can express the decay amplitude along with the new physics contributions as
\bea\label{NP Amp VLDQ}
\mathcal{A}_{B_s^0 \to K^+K^-} &=&A_{K\bar{K}}\big[\delta_{\mathit{p}u} \alpha_1 + \alpha _4^{\mathit{p}}+\alpha_{4,EW}^{\mathit{p}}+ \beta _3^{\mathit{p}}+  \beta _4^{\mathit{p}}-\frac{1}{2} \beta _{3, EW}^{\mathit{p}}-\frac{1}{2}\beta_{4,EW}^{\mathit{p}}\big] \nn\\
& +& B_{\bar{K}K}\big[\delta _{\mathit{pu}}b_1+b_4^p+b_{4, EW}^{\mathit{p}}\big] \nn\\
&-&Q_{bs} \Big [ A_{K\bar{K}}\big (\tilde{ \alpha} _4^{\mathit{p}}+\tilde{\alpha}_{4,EW}^{\mathit{p}}+ \tilde{\beta} _3^{\mathit{p}}-\frac{1}{2} \tilde{\beta} _{3, EW}^{\mathit{p}} \big) \Big ]\,.
\eea
Here,  $\tilde{\alpha}^p$ and $\tilde{\beta}^p$ provides the dominant contributions  to NP amplitude which contain all the above three new couplings as given in the equation \ref{Br VLDQ}\,. Symbolically,  the full amplitude   can be written as
\bea
\mathcal{A}&=& \xi_u \mathcal{A}_u +\xi_c \mathcal{A}_c-Q_{bs} \mathcal{A}_{NP} \nn\\
&=& \xi_c A_c\Big[1+\varrho a^\prime e^{\mathit{i}(\delta _1 ^\prime-\gamma)}-\varrho ^\prime  b^\prime  e ^{\mathit{i}(\delta _2 ^\prime + \phi _s)}\Big]\,,
\eea
where 
\begin{align}
a^\prime=|\frac{\xi _u}{\xi _c}|, \hspace{3mm} b^\prime=|\frac{Q_{bs}}{\xi _c}|, \hspace{3mm}
\varrho=|\frac{A_u}{A_c}|,  \hspace{3mm} \varrho ^ \prime =|\frac{A_{NP}}{A_c}|.
\end{align}
Here, $\gamma$ is the weak phase of $V_{ub}, \phi _s$ is the weak phase of $Q_{bs}$ and $\delta _1^\prime (\delta _2 ^ \prime)$ is the relative strong phase between $A_u$ and $A_c (A_{NP}$ and $A_c)$. From the parametrized amplitude, the CP-averaged branching ratio  can be written as
\bea\label{NP BR}
\lbrace\mathcal{BR}\rbrace &=&\frac{\tau _{B_s} p_c}{8 \pi m_{B_s}^2} |\xi _c A_c|^2\bigg[\mathcal{G}^\prime+2\varrho a^\prime\cos \delta _1 ^\prime \cos \gamma -2 \varrho ^\prime b\cos \delta _2 ^\prime \cos \phi _s  \nn \\&& -2 \varrho \varrho ^\prime a^\prime b^\prime \cos(\delta _1 -\delta _2 ^\prime) \cos (\gamma + \phi _s)\bigg] \,,     
\eea
On the other hand, the direct CP asymmetry can be written as
\bea
\mathcal{C}_{KK}=
-\frac{2\big[\varrho a^\prime \sin \delta _1^\prime \sin \gamma + \varrho^\prime b^\prime \sin \delta _2 ^\prime \sin \phi _s +\varrho \varrho ^\prime a b^\prime \sin (\delta _2 ^\prime-\delta _1)\sin( \gamma +\phi _s)\big]}{\big[ \mathcal{G}^\prime+2\big(\varrho a^\prime \cos \delta _1 ^\prime \cos \gamma - 2 \varrho ^\prime b^\prime \cos \phi _s \delta _2 ^\prime-2 \varrho \varrho ^\prime a^\prime b ^\prime\cos (\gamma + \phi _s)\cos (\delta _2 ^\prime -\delta _1 ^\prime)\big)\big]}\,,
\eea
One can obtain the mixing induced CP asymmetry parameter as
\bea
\mathcal{S}_{KK}=\frac{\mathcal{M} ^\prime}{\mathcal{G} ^\prime+2 \varrho a^ \prime \cos \delta _1^\prime \cos \gamma -2\varrho_2b^\prime  \cos \delta _2 ^\prime \cos \phi _s-2 \varrho _1 \varrho _2 a^\prime b^\prime \cos (\delta_1 ^\prime -\delta _2 ^\prime) \cos(\gamma + \phi _s)},
\eea
where $\mathcal{G} ^\prime=1+(\varrho  a^\prime)^2+ (\varrho ^\prime b^\prime)^2$ and
\begin{align}
\mathcal{M}=&\sin 2 \beta _s +2 r a \cos \delta _1 \sin(2 \beta _s +\gamma)-2 r^\prime  b \cos \delta _2 \sin(2 \beta_s-\phi _s)+(r a)^2\sin (2 \beta_s +2 \gamma) \\ 
+& (r^\prime b)^2\sin(2 \beta_s -2 \phi _s)-2 r r^\prime a b \cos(\delta -\delta ^\prime) \sin(2 \beta_s + \gamma -\phi _s).
\end{align}

\begin{figure}[htb]
\centering
\includegraphics[scale=0.85]{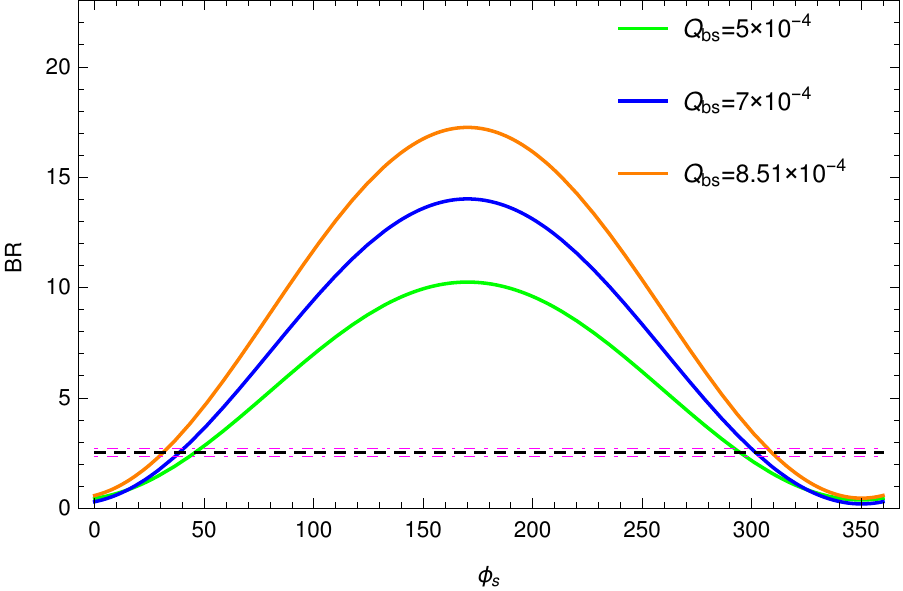} 
\quad
\includegraphics[scale=0.85]{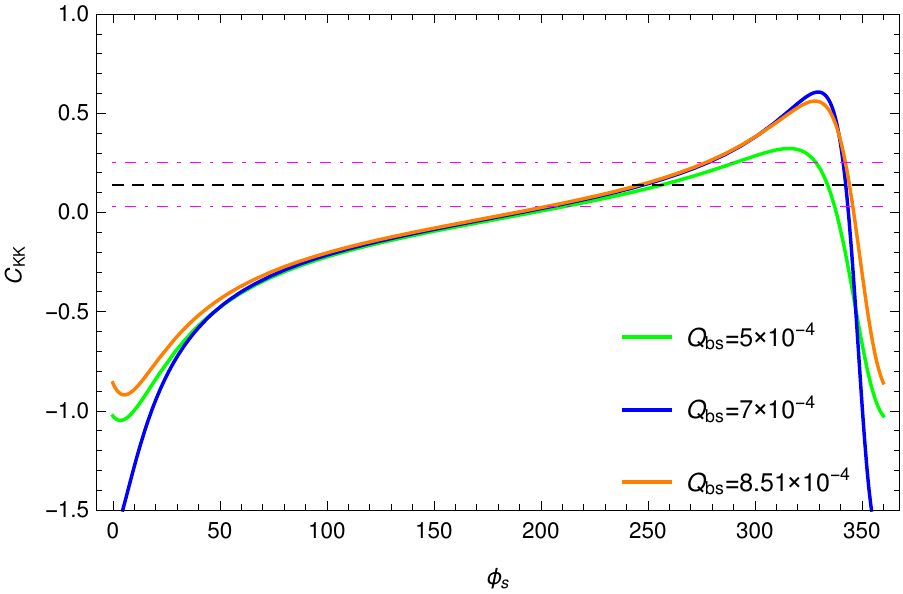} 
\quad
\includegraphics[scale=0.85]{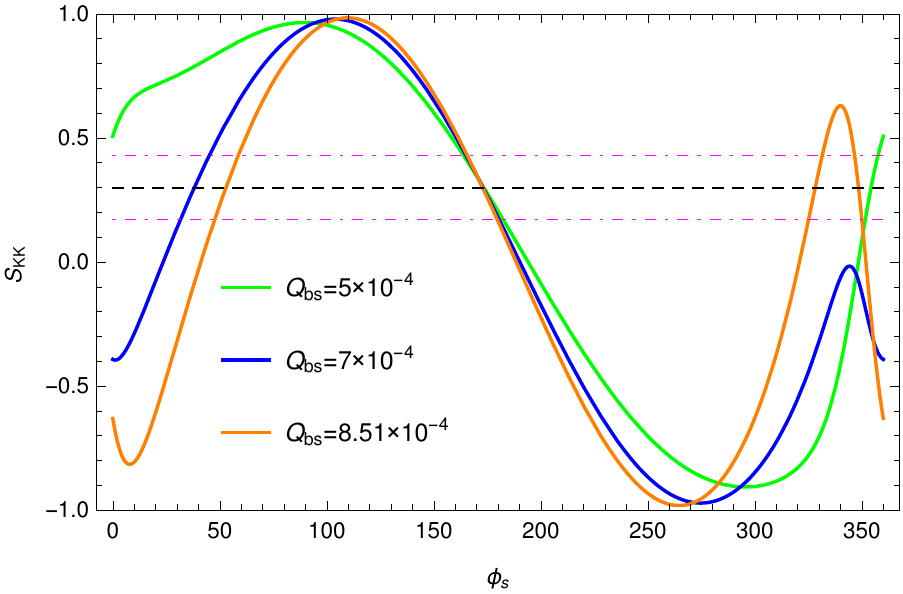} 
\caption{Variation of CP-averaged branching ratio (in the units of $10^{-5})$ (top-left panel), direct CP asymmetry(top-right panel) and mixing induced CP asymmetry(bottom panel) with the new weak phase $\phi _s$ for different $|U_{bs}|$ entries. The horizontal solid line (black color) represent the central experimental value where as the dotted lines (magenta color) denote the 1$\sigma$ error limit.}
\label{Fig:result5}
\end{figure}

\begin{figure}[htb]
\centering
\includegraphics[scale=0.6]{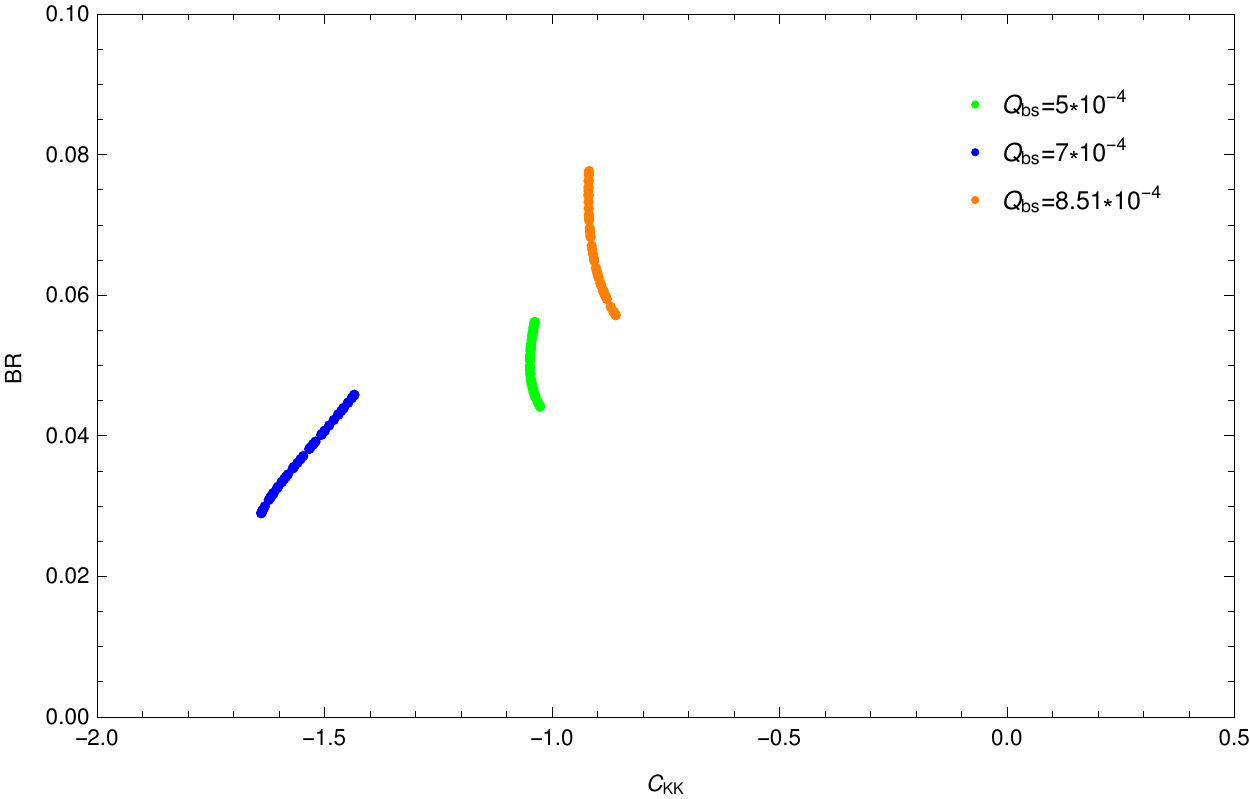} 
\quad
\includegraphics[scale=0.6]{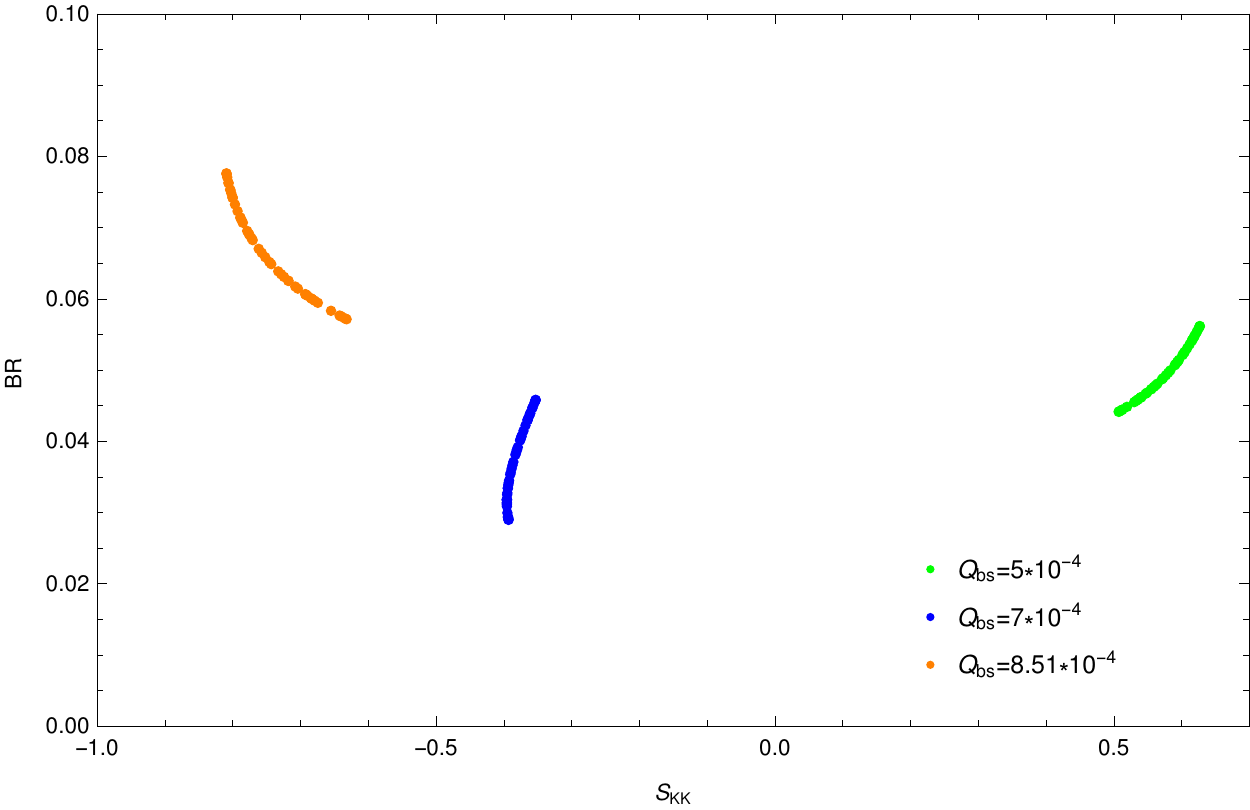} 
\quad
\includegraphics[scale=0.6]{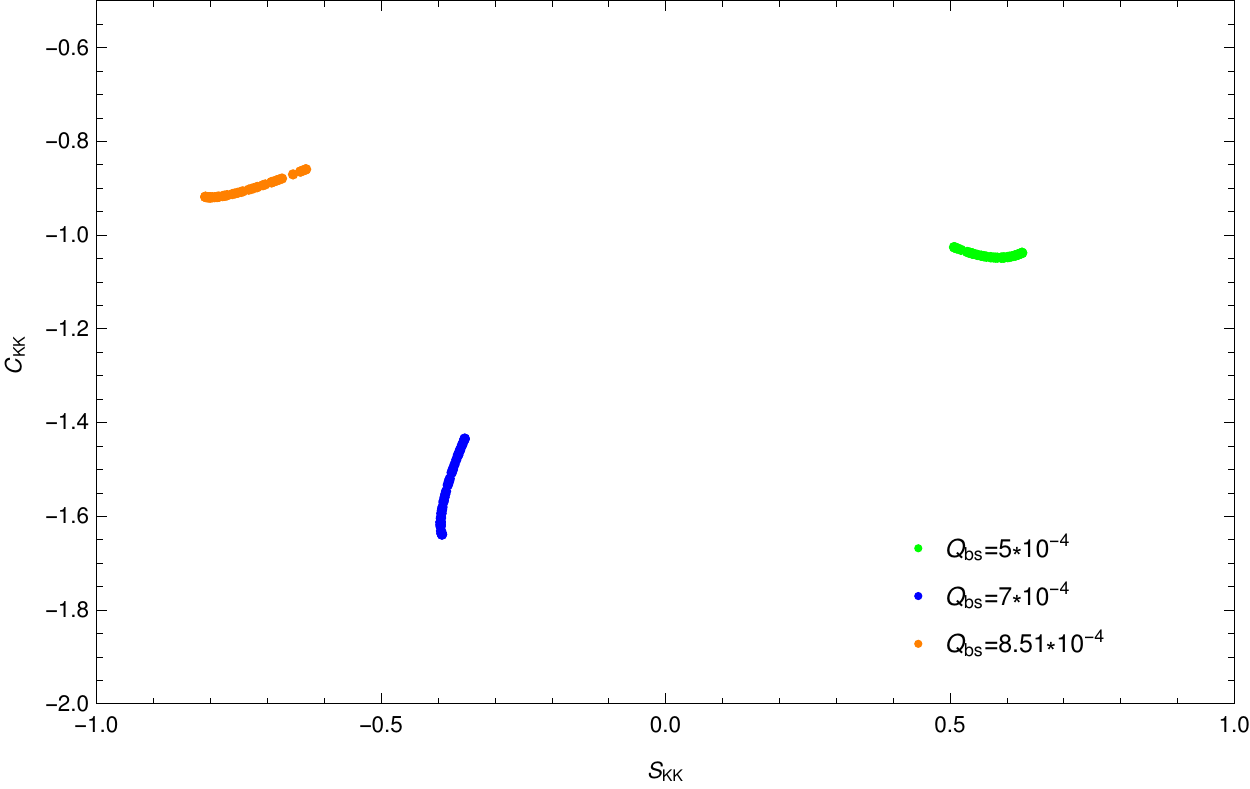} 
\caption{Correlation plots of CP-averaged branching ratio (in the units of $10^{-4})$ with direct CP asymmetry (top-left panel) and mixing induced CP asymmetry (top-right panel) (in \%), and direct CP asymmetry with mixing induced CP asymmetry (bottom panel).}
\label{Fig:result6}
\end{figure}

In FIG. \ref{Fig:result5}\,, we present the variation of CP-averaged branching ratios (top-left panel), direct (top-right panel) and mixing induced CP asymmetry (bottom panel) with respect to the weak phase $\phi_s$ for three benchmark $Q_{bs}$ values. Here, green,blue and orange colors represent the predictions obtained by using  $Q_{bs}$= $5\times 10^{-4}$, $7\times 10^{-4}$ and $8.51\times 10^{-4}$, respectively.  For $\phi$ below $50^\circ$ (above $290^\circ$) and all the discussed benchmark points of $Q_{bs}$, the predicted branching of $B_s \to KK$  are accommodating within $1 \sigma$ range of experimental central value. We also notice that, the CP violating parameters are also explained within $1\sigma$ data. The correlation plots among all the  observables are shown in  FIG. \ref{Fig:result6} for the  same  inputs of $Q_{bs}$. 
The predicted results of branching ratio, CP violating observables for different values of $Q_{bs}$ and $\phi _s$  are shown  in the bottom section of Table \ref{Tab:result}\,.

\begin{table}
\centering
\caption{Predicted values of CP-averaged branching ratio and CP violating observables for different  benchmark values of $U_{bs}~(Q_{bs})$ and $\phi _s$ parameters  in the $Z^\prime$~(VLDQ) model.}\label{Tab:result}
\begin{tabular}{|l|l|l|l|l|l|l|}
\hline

Model~&~$U_{bs}(Q_{bs})$ ~& ~$\phi _s$ ~ &~ $\mathcal{BR}$ ~ &~  $C_{KK}$~  &~  $S_{KK}$~  \\

\hline \hline

\multirow{3}{*}{} $Z^\prime$ Model~&~$0.1$ & $ \makecell{0^\circ\\180^\circ}$ & $\makecell{54.06\times 10^{-6}\\ 21.76\times 10^{-6} }$   &  $\makecell{-0.04\\-0.25 }$  &  $\makecell{0.24\\0.42 }$  \\
\cline{2-6}

&~$0.3$ &  $ \makecell{0^\circ\\180^\circ}$  & $ \makecell{11.46\times 10^{-5} \\ 17.75\times 10^{-6}}$ & $\makecell{0.007\\ -0.49 }$ &   $\makecell{0.15\\ 0.32 }$\\ 

\cline{2-6}
&~$0.5$~ & $ \makecell{0^\circ\\180^\circ}$  & $\makecell{20.35\times 10^{-5}\\ 4.2\times 10^{-5} }$ & $\makecell{0.02\\ -0.28 }$ &  $\makecell{0.11 \\ 0.07}$  \\ 
\hline
VLDQ Model~&~$5\times 10^{-4}$ & $ \makecell{0 ^{\circ}\\180^{\circ}}$ & $\makecell{4.41\times 10^{-6}\\ 10.17\times 10^{-5}}$   &  $\makecell{-1.14\\ -0.02 }$  &  $\makecell{0.50\\0.20 }$  \\  
 \cline{2-6}
&~$7\times 10^{-4}$ &   $ \makecell{0^{\circ}\\180^{\circ}}$  & $ \makecell{2.89\times 10^{-6} \\ 13.91\times 10^{-5}}$ & $\makecell{-1.90\\ -0.01 }$ &   $\makecell{0.39\\ 0.17 }$\\ 
\cline{2-6}
&~$8.51\times 10^{-4}$ & $ \makecell{0^{\circ}\\180^{\circ}}$  & $\makecell{5.7\times 10^{-6} \\ 17.13\times 10^{-5}}$ & $\makecell{-1.02 \\ -0.012 }$ &  $\makecell{-0.63 \\0.164 }$  \\ 
\hline
\end{tabular}

\end{table}


\section{Conclusion}
We have investigated the observables of $B_s^0 \to K^+ K^-$, a penguin induced decay mode occurring at quark level transition $b \to s$, both in standard model as well as beyond the SM scenarios.
In the new physics scenario, we consider both the  $Z ^\prime$ and  vector-like down quark  model,  where the consequence of the former one is nothing but the minimal extension of SM having $U(1)^ \prime$ gauge group added to it and the later one  provides the interaction of $Z$ mediated FCNC  at the tree level. 
We have constrained the NP parameter associated with ``$Z^{(\prime)}-b-s$'' interactions from the branching ratios of all leptonic $B_s$ decay modes and mainly checked whether the new physics coupling has impact on the physical observables of non-leptonic $B_s\to K^+K^-$ decay mode.
We have found that the CP-averaged branching ratios  have deviated  significantly  from the SM results  for sizable new physics parameters $\rm U_{bs}(\rm Q_{bs})$ of $Z^\prime \rm(VLDQ)$ model. Furthermore, the CP violating parameters such as direct and mixing-induced also have profound deviation  in the presence of new physics. 
To conclude, we have noticed that the observables of $B_s \to KK$ mode can be explained in  both  $Z^ \prime$ and VLDQ models. 

\acknowledgments 

MKM would like to thank to Department of Science and Technology(DST)- Inspire Fellowship division, Government of India for financial support through ID No- IF160303.  MKM  would like to acknowledge Prof. Anjan Giri for his support and
useful discussion.
\bibliography{Bs-KK}



\end{document}